\documentclass[lettersize,journal]{IEEEtran}
\usepackage{amsmath}
\usepackage{amsthm}
\usepackage{amsfonts}
\usepackage{amssymb}
\usepackage{url}
\usepackage{framed}
\usepackage{graphicx} 
\usepackage{caption}
\usepackage{color}

\begin{document}


\title{Microfluidic Pulse Shaping Methods for \\ Molecular Communications}
\author{Maryam Kahvazi Zadeh, Iman Mokari Bolhassan,
        and Murat Kuscu
       \thanks{The authors are with the Nano/Bio/Physical Information and Communications Laboratory (CALICO Lab), Department of Electrical and Electronics Engineering, Koç University, Istanbul, Turkey (Corresponding authors' e-mail: \{mzadeh22, mkuscu\}@ku.edu.tr).}
	   \thanks{This work was supported in part by the European
Union’s Horizon 2020 Research and Innovation Programme through the Marie Skłodowska-Curie Individual Fellowship under Grant Agreement 101028935, and by The Scientific and Technological Research Council of Turkey (TUBITAK) under Grant \#120E301.}}

\maketitle

\begin{abstract}

Molecular Communication (MC) is a bio-inspired communication modality that utilizes chemical signals in the form of molecules to exchange information between spatially separated entities. Pulse shaping is an important process in all communication systems, as it modifies the waveform of transmitted signals to match the characteristics of the communication channel for reliable and high-speed information transfer. In MC systems, the unconventional architectures of components, such as transmitters and receivers, and the complex, nonlinear, and time-varying nature of MC channels make pulse shaping even more important. While several pulse shaping methods have been theoretically proposed for MC, their practicality and performance are still uncertain. Moreover, the majority of recently proposed experimental MC testbeds that rely on microfluidics technology lack the incorporation of programmable pulse shaping methods, which hinders the accurate evaluation of MC techniques in practical settings. To address the challenges associated with pulse shaping in microfluidic MC systems, we provide a comprehensive overview of practical microfluidic chemical waveform generation techniques that have been experimentally validated and whose architectures can inform the design of pulse shaping methods for microfluidic MC systems and testbeds. These techniques include those based on hydrodynamic and acoustofluidic force fields, as well as electrochemical reactions. We also discuss the fundamental working mechanisms and system architectures of these techniques, and compare their performances in terms of spatiotemporal resolution, selectivity, system complexity, and other performance metrics relevant to MC applications, as well as their feasibility for practical MC applications.
\end{abstract}


\section{Introduction}\label{Int}
\IEEEPARstart{T}he development of unconventional communication systems that can process and exchange information via molecules and chemical reactions is a rapidly growing research field, holding the promise of revolutionizing computing and communications, and extending our connectivity to micro/nanoscale and biological devices and entities. Molecular Communications (MC) is a bio-inspired way of communication that uses molecules for encoding, transmission, and reception of information, and is already ubiquitous among living cells \cite{akyildiz2019moving, kuscu2019transmitter}. MC is a highly interdisciplinary research area at the intersection of information and communication technologies (ICT), micro/nanotechnology, and biotechnology. A variety of paradigm-shifting applications can be achieved with the help of MC systems, particularly in the biomedical field. The envisioned applications mostly concern the early diagnosis and treatment of diseases, such as continuous health monitoring, targeted drug delivery in nanomedicine, artificial organs, and lab-on-a-chip, in which the use of electromagnetic signals is either not desirable due to biocompatibility concerns or not feasible due to physical constraints \cite{felicetti2016applications, akan2016fundamentals, bi2021survey, barros2021molecular, kuscu2021internet}. Furthermore, MC can be deployed in industrial settings for monitoring chemical reactors and nanoscale manufacturing, as well as for larger-scale practices, such as monitoring the emissions of pollutants and the safe and clean transportation of oil \cite{akyildiz2011nanonetworks, haselmayr2019integration}.

\begin{figure}[t!]
    \centering
    \includegraphics[width=\linewidth]{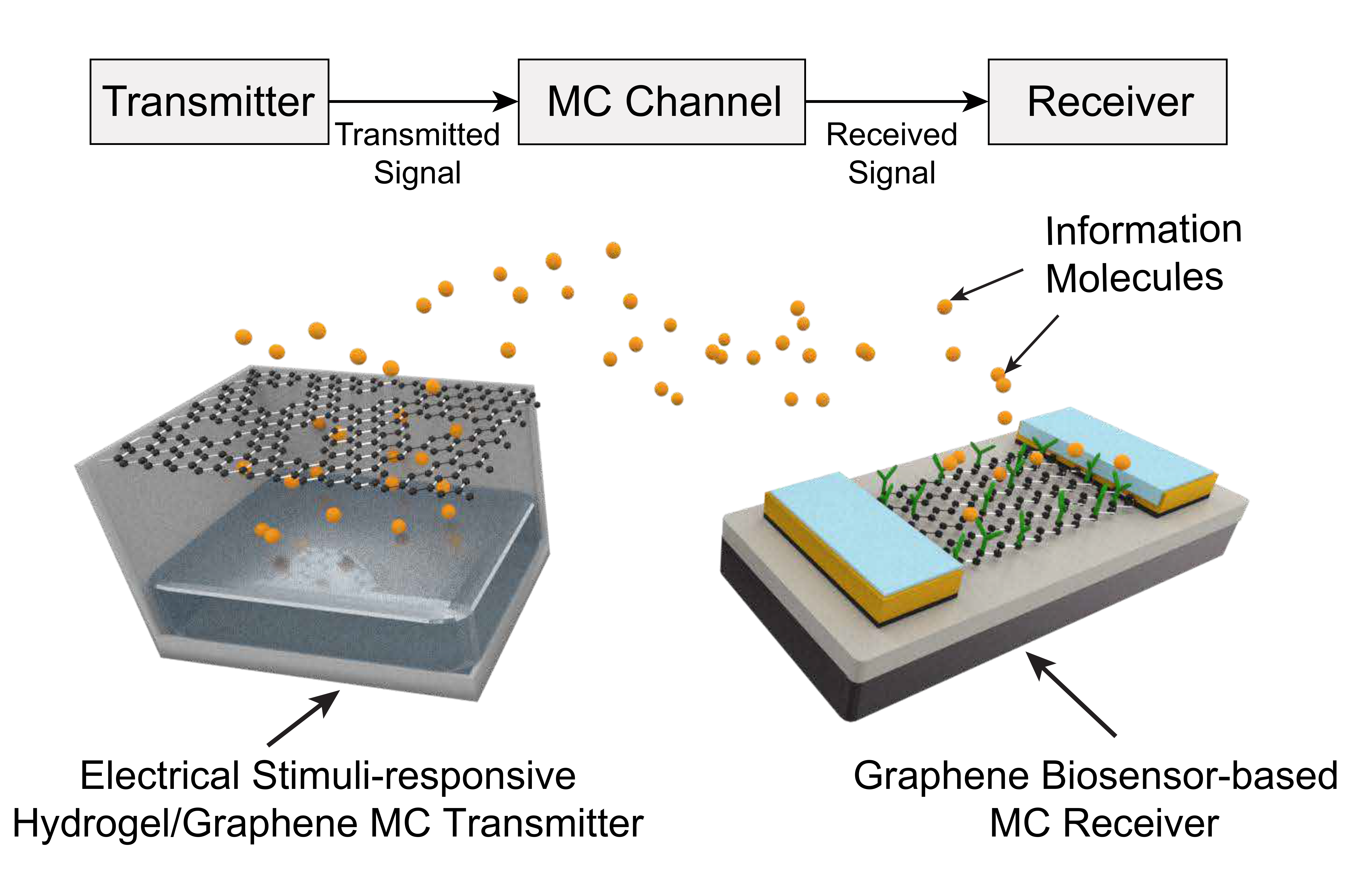}
        \caption{A conceptual drawing of an MC system with nanomaterial-based transmitter and receiver \cite{kuscu2019transmitter}.} 
    \label{fig:molecularcommunication}
\end{figure}

Unlike conventional communication systems, which typically use electromagnetic signals as communication carriers, MC uses chemical signals, as conceptually illustrated in Fig. \ref{fig:molecularcommunication}. Accordingly, a transmitter in an MC system releases molecules into a fluidic medium, where the molecules travel through diffusion and/or drift, and a portion of them manages to reach an MC receiver. Typically, information molecules cause a certain reaction at the receiver, through which the receiver detects and decodes the transmitted information encoded into a distinguishable property of the molecules, such as their concentration, type, or release time from the transmitter \cite{hiyama2006molecular, farsad2016comprehensive, kuscu2022detection}. 

While the theoretical aspects of MC have been significantly researched, its practical aspects concerning the design and prototyping of MC components and systems are yet to be fully addressed \cite{kuscu2019transmitter}. This is partly due to the highly interdisciplinary technical knowledge and tools required for constructing such multi-physics and multi-scale systems \cite{deng2017microfluidic,balasubramaniam2013review}. In spite of all these issues, a number of experimental studies were performed on microfluidic testbeds that can replicate the flow conditions of biological environments typically considered for MC applications \cite{kuscu2021fabrication, amerizadeh2021bacterial}.
However, there are important challenges, including the lack of high-resolution control of the spatiotemporal distribution of molecules inside the microfluidic channels, that hinder the accurate experimental validation of the theoretical channel models, as well as the tests and the refinement of the developed MC modulation and detection techniques \cite{kuscu2020nano, kuscu2019transmitter, farsad2016comprehensive}.

\begin{figure}[t!]
    \centering
    \includegraphics[width=\linewidth]{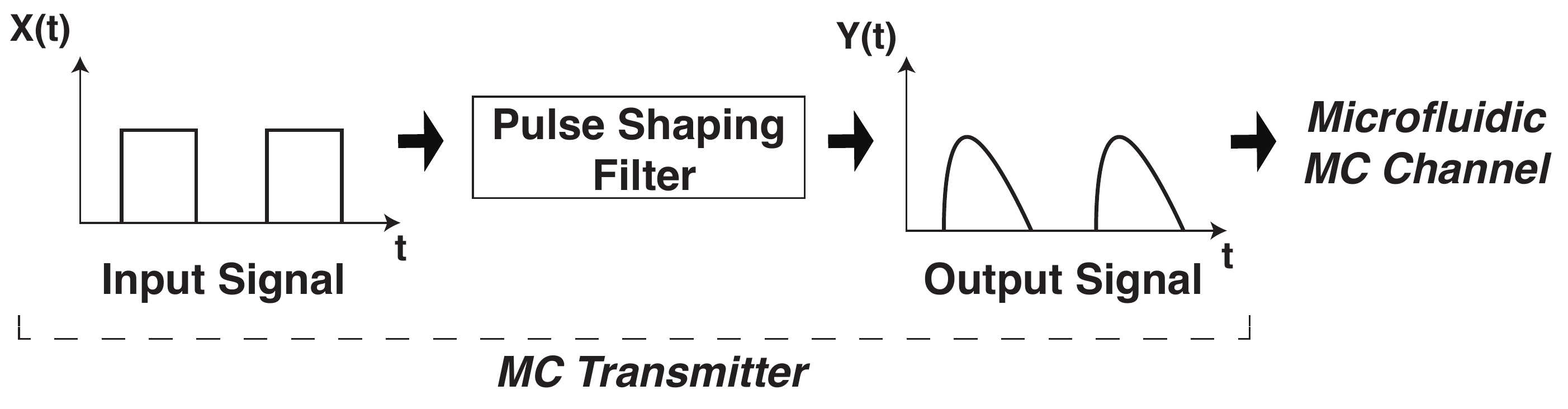}
        \caption{The overall scheme of the pulse shaping in a microfluidic MC system.} 
    \label{fig:pulse_shaping}
\end{figure}

Pulse shaping is of critical importance in all communication systems for enabling accurate and high-rate information transfer by adjusting the waveform of the signals to be transmitted (see Fig. \ref{fig:pulse_shaping}) in light of the communication channel's characteristics, e.g., bandwidth, thereby, minimizing the effects of noise and intersymbol interference (ISI), and eventually improving the received signal strength, or equivalently, its signal-to-interference-plus-noise ratio (SINR) \cite{proakis2007fundamentals}.  
As the fluidic MC channel manifests peculiar properties, which are quite different than its conventional electromagnetic counterparts, such as the low-pass characteristics due to the slow diffusion of information carriers (i.e., molecules) the importance of pulse shaping is more pronounced in MC \cite{akan2016fundamentals, kuscu2018modeling}. Moreover, the unique micro/nanoscale architectures of the transmitter and the receiver components of the MC system can impose further challenges in the transmission and reception of the messages encoded into molecules \cite{kuscu2019transmitter, kuscu2016physical, rezaei2022molecular}. Furthermore, the discrete nature of information carriers necessitates rethinking the conventional pulse shaping techniques that have been typically developed for continuous information carriers, e.g., EM waves. 

The radically different nature of communication channels, information carriers, transmitter, and receiver architectures in MC also create unique opportunities for pulse shaping, some of which have already been explored. For example, different diffusion characteristics of molecules of varying sizes were exploited to form numerous MC signal waveforms consisting of multiple types of molecules with well-defined mixtures \cite{wicke2022pulse}. Additionally, it was shown that the timing and the duration of chemical reactions of molecules with certain enzymes can be tuned to form well-defined MC pulses inside microfluidic channels \cite{deng2017microfluidic, bi2020chemical}. Moreover, there were efforts to control the physical characteristics of the boundaries of the propagation medium, i.e., the communication channel, to form arbitrary MC pulse shapes \cite{bicen2013system, kuran2013tunnel}. Although, these approaches demonstrated the enhancement of MC performance with pulse shaping, the practicality of the employed techniques and their actual performances are questionable, as they have not been experimentally validated yet.   

Generating well-defined and tunable chemical waveforms and concentration gradients in microfluidic systems has also attracted considerable attention in various research fields other than MC. For example, in biological research, the need for dynamically controlling the molecular composition of the extracellular matrix (ECM) of cultured biological cells and tissues in microfluidic lab-on-a-chip systems motivated the development of arbitrary concentration waveform generators \cite{chen2020review}. The proposed techniques utilize different external force fields, such as acoustical or hydrodynamic force fields, or electrochemical reactions, to program the spatiotemporal distribution of molecules inside microfluidic channels with high resolution, and generate well-defined concentration waveforms that can take arbitrary shapes \cite{abadie2019electrochemical, chen2010hydrodynamic, huang2018sharp, ahmed2013tunable}. Depending on the nature of the external control, the system architectures of the waveform generators show a large variety in terms of complexity. Their performances in terms of selectivity, repeatability, and spatiotemporal resolution of the waveforms they generate also differ significantly. Nevertheless, the practicality and performances of these techniques have been validated through extensive experimental studies, and thus, they are promising as externally-controlled pulse shaping techniques for use in practical microfluidic MC systems and testbeds. 

In this review article, our objective is to thoroughly analyze microfluidic chemical waveform generation methods that can inform the design of pulse shaping techniques in practical microfluidic MC testbeds and systems. We aim to provide a comprehensive overview of these techniques, including their working mechanisms, system architectures, and additional components required for external control. We also conduct a comparative analysis of these methods in terms of their spatiotemporal resolution and control, selectivity, and system complexity, and evaluate their feasibility for a range of practical MC applications. We believe that this review will be a valuable resource for researchers seeking to innovate in the field of MC, as it will help to bridge the gap between theory and practice by providing a detailed examination of these methods and their suitability for use in microfluidic MC testbeds and systems. 

The remainder of this paper is organized as follows: In Section \ref{sec:pulse_shaping}, we explore the significance of pulse shaping in MC and review the various methods that have been proposed for this purpose in the MC literature. In the following section, we delve into microfluidic chemical waveform generation techniques that enable the production of programmable, dynamic chemical signals, examining their underlying mechanisms, system architectures, and additional components required for external control. Next, in Section 4, we comprehensively evaluate the performance of each pulse shaping technique in terms of spatiotemporal resolution and control, selectivity, repeatability, control over propagation, and system complexity. We also discuss their feasibility for integration into practical MC applications in Section \ref{sec:discussion-feasibility}. Finally, we conclude the paper in Section \ref{sec:conclusion}.

\section{Pulse Shaping in Molecular Communications}
\label{sec:pulse_shaping}
The characteristics of each component in the MC system, e.g., transmitter, channel, and receiver, exhibit additional communication challenges which highlight the need for pulse shaping in MC systems. For instance, at the transmitter, due to the limited molecule generation and storage capacity, it is necessary to design molecule-efficient pulses in order to transmit information molecules effectively. In addition, the low-pass characteristic of the MC channel, which typically relies on diffusion, leads to significant spatial dispersion of the molecular signals as they propagate in the microfluidic channel \cite{wicke2022pulse}. This may result in substantial ISI that hampers the achievable data rates in MC channels \cite{kuscu2019transmitter}. The ISI effect is, also, compounded by the fact that practical MC receivers, such as nanobiosensors and engineered bacteria, typically employ ligand receptors, which interact selectively with the transmitted information molecules, i.e., ligands, at finite reaction rates, adding to the ISI observed at the receiver. Furthermore, MC receivers with ligand receptors typically manifest nonlinear responses to the incoming ligand concentrations, and therefore, can be severely affected by the saturation of the receptors when the received signals in terms of ligand concentrations are not optimized. Likewise, in a ligand-receptor-based MC receiver, the accuracy of the detection is dependent on the received ligand concentration, such that when the receptors are only very shortly exposed to the ligand concentration, because of the finite reaction rates, there may not be a sufficient number of bound receptors for detection \cite{kuscu2019channel}. 

To overcome the unique challenges of end-to-end MC channels, transmission pulses can be optimized according to the modulation technique adopted by the transmitter. This optimization per se is not sufficient, as we also need practical systems that can generate the optimized MC pulse shapes to overcome the challenges associated with this system \cite{wicke2022pulse, kang2011pulse, kuran2013tunnel}. There have been a few proposed pulse shaping methods considered for MC in the literature in order to address some of these challenges, which are reviewed in the remainder of this section.

\subsection{Pulse Shaping based on Molecular Propagation Characteristics}
\label{sec:pulse_shaping_propagation}
Diversity in the propagation characteristics of different types of molecules can be used to shape the transmitted and received molecular signals in MC systems. For example, Wicke et al. \cite{wicke2022pulse} proposed an MC pulse shaping method exploiting different diffusion coefficients of molecules that vary in size. Accordingly, at the transmitter, they tuned and optimized the molecule size mixture to guarantee both a minimum signal level and a maximum signal level. A minimum signal level was considered for detecting the presence of the signal, and a maximum signal level was considered for detecting the absence of the signal within a predetermined detection window. With this method, suitable molecule sizes can be selected optimally for any required detection duration in comparison to mixing all molecules together. Moreover, their optimization framework can be extended to the cases where the channel impulse responses (CIR) of individual types of molecules depend on molecule properties other than their size or can be determined only empirically through experiments or simulations of complex and reactive environments.

\begin{figure}[t]
    \centering
    \includegraphics[width=0.48\textwidth]{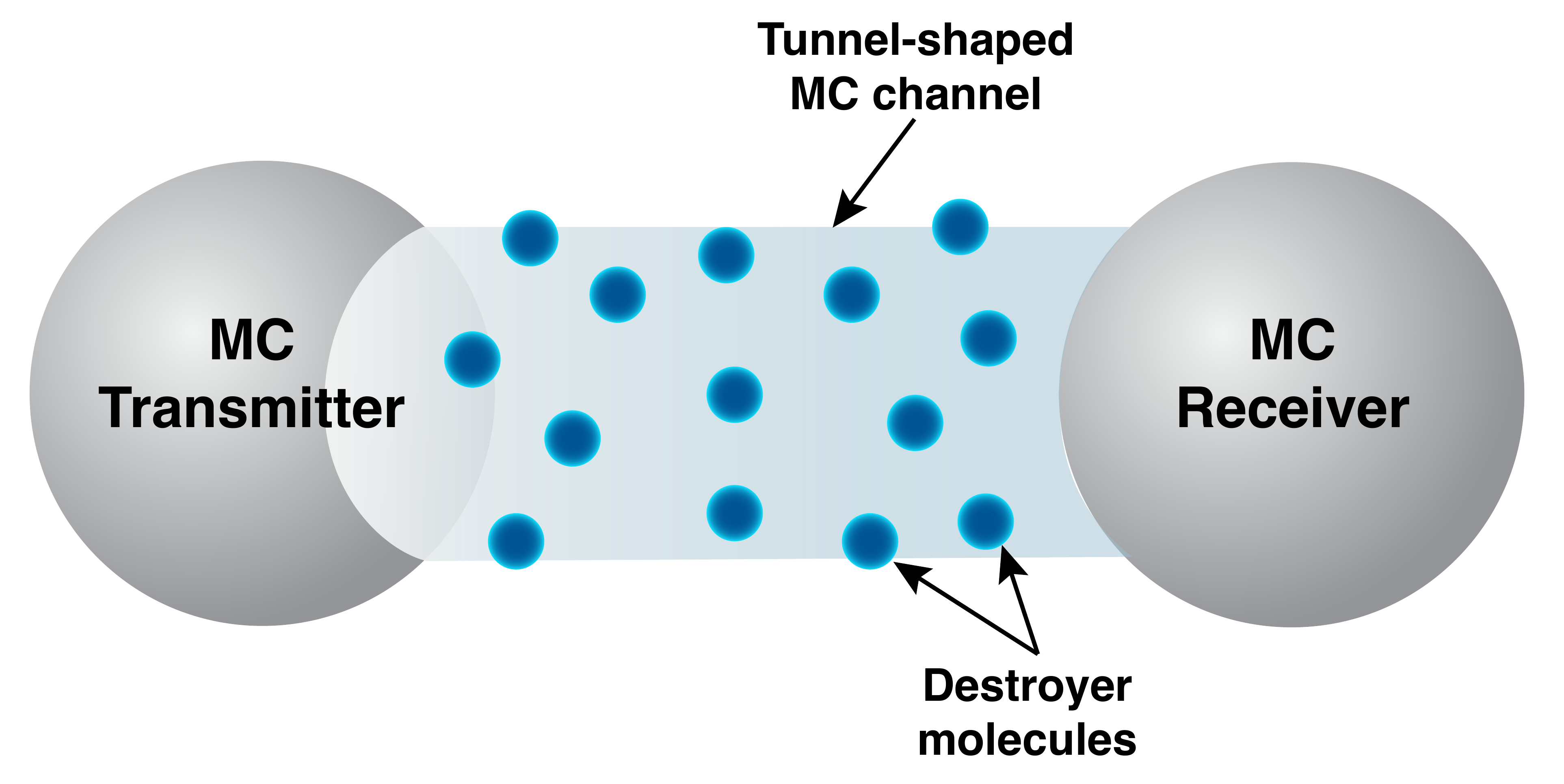}
        \caption{A cylindrical tunnel shape environment for diffusion-based MC system \cite{kuran2013tunnel}.}
    \label{fig:tunnelbased}
\end{figure}

\subsection{Pulse Shaping Based on Chemical Reactions}
\label{sec:pulse_shaping_chemical_reactions}
By tuning the rate and the duration of chemical reactions, and the concentration of reactants (i.e., input molecules) in a controlled microfluidic system, the time-domain concentration profile of the reaction products (i.e., output molecules) can be patterned for encoding information. In \cite{deng2017microfluidic}, based on a triggering chemical input, the authors proposed a pulse generator for MC which produces predefined molecular concentration pulses that are shaped in the transmitter part of the system and then transmitted into and propagated through a microfluidic channel. The proposed design consists of a microfluidic system with standard and reproducible components, whose geometry and flow conditions effectively help determine the resulting pulse shape. Pulse generation in the proposed method is inspired by the motifs observed in cells' gene regulatory networks and compatible with the molecular pulse-modulation techniques investigated in the MC literature  \cite{pehlivanoglu2017modulation, huang2019spatial, alon2007network}. The study also presented a methodology with a modeling and analysis framework for diversifying the generated pulse shapes with the integration of additional microfluidic components inspired by biochemical processes. 

The same group also developed a microfluidic receiver in \cite{bi2020chemical} using thresholding and amplification reactions to achieve the demodulation of received signals into rectangular output signals in the form of concentration of molecules. The microfluidic components were characterized as part of an analytical study to determine how several design parameters such as the concentration of molecules, fluid viscosity, diffusion coefficient, and the length of injection time of the sample into the system influence the generated pulse and the demodulated signal. To optimize the transmitter design, the authors proposed a reaction channel length optimization flow to control the maximum output pulse concentration at the transmitter. They then derived a time gap constraint between two consecutive input signals to ensure the continuous transmission of non-distorted and identical-shaped pulses.\\

\subsection{Tunnel-based Pulse Shaping}
\label{sec:pulse_shaping_tunnel}
 To reduce the variance of the receiver-hit time distribution of transmitted molecules in diffusion-based MC, which causes high propagation delays and ISI, the authors in \cite{kuran2013tunnel} proposed developing a tunnel-based pulse shaping method for MC. This method relies on the use of particular types of molecules, called destroyer molecules. These molecules clean the MC channel from the information molecules remaining from the previous transmissions, and thus, control the shape of the received signals. In this study, the authors suggested a system with an MC channel surrounded by AChE-like destroyer molecules to construct a tunnel between the transmitter and the receiver, as illustrated in Fig. \ref{fig:tunnelbased}. The use of destroyer molecules is inspired by the Acetylcholine (ACh) - Acetylcholinesterase (AChE) interaction observed in the Neuromuscular Junctions (NMJs) of living organisms. NMJs connect muscle cells to nerve cells via an intercellular gap called the synaptic cleft. AChE molecules destroy messenger ACh molecules after sending a contraction signal to return muscle cells to their resting state. It is important to note that the proposed tunnel-based pulse shaping method operates in a distinct manner compared to other conventional and MC pulse techniques such that pulse shaping through chemical reactions in this method is realized within the channel through the utilization of destroyer molecules.
 
 In this way, the received signal variance is reduced and the communication channel is cleaned for the next signal. Hence, the proposed tunnel-based pulse shaping method can control the spread and hence the shape of the molecular signals with the help of destroyer molecules and the guidance of the tunnel geometry, in order to reduce the received signal variance and the ISI, thereby significantly increasing the channel capacity. 

 \section{Microfluidic Pulse Shaping Methods}
\label{sec:microfluidic_pulse_shaping}

Microfluidic chemical waveform generation techniques that can produce dynamic chemical signals with programmable shapes in microfluidic channels have recently attracted significant interest in many branches of biological research concerning the monitoring and manipulation of physiological processes, such as immune response, development, embryogenesis, and cancer metastasis \cite{perrodin2020electrochemical}. One prominent application of chemical waveform generation in this area is the dynamic control of the chemical composition in the ECM of living cells cultured in microfluidic platforms, ensuring high spatiotemporal resolution \cite{ahmed2014acoustofluidic}.

After a brief review of the operating mechanisms and the key properties of several microfluidic chemical waveform generation techniques in this section, a detailed comparison of them is provided in the succeeding sections to comprehensively evaluate their potential for use in microfluidic MC testbeds and systems as pulse shaping techniques.

\subsection{Hydrodynamic Methods}
\label{sec:hydrodynamic}
Chemical concentration waveforms in microfluidic channels can be formed through hydrodynamic control, such that the modulation of the flow velocity and direction can be exploited to tune the injection rate of molecules into the microfluidic channel and shape their concentration waveforms propagating inside the channel. 

\begin{figure}[t!]
    \centering
    \includegraphics[width=0.48\textwidth]{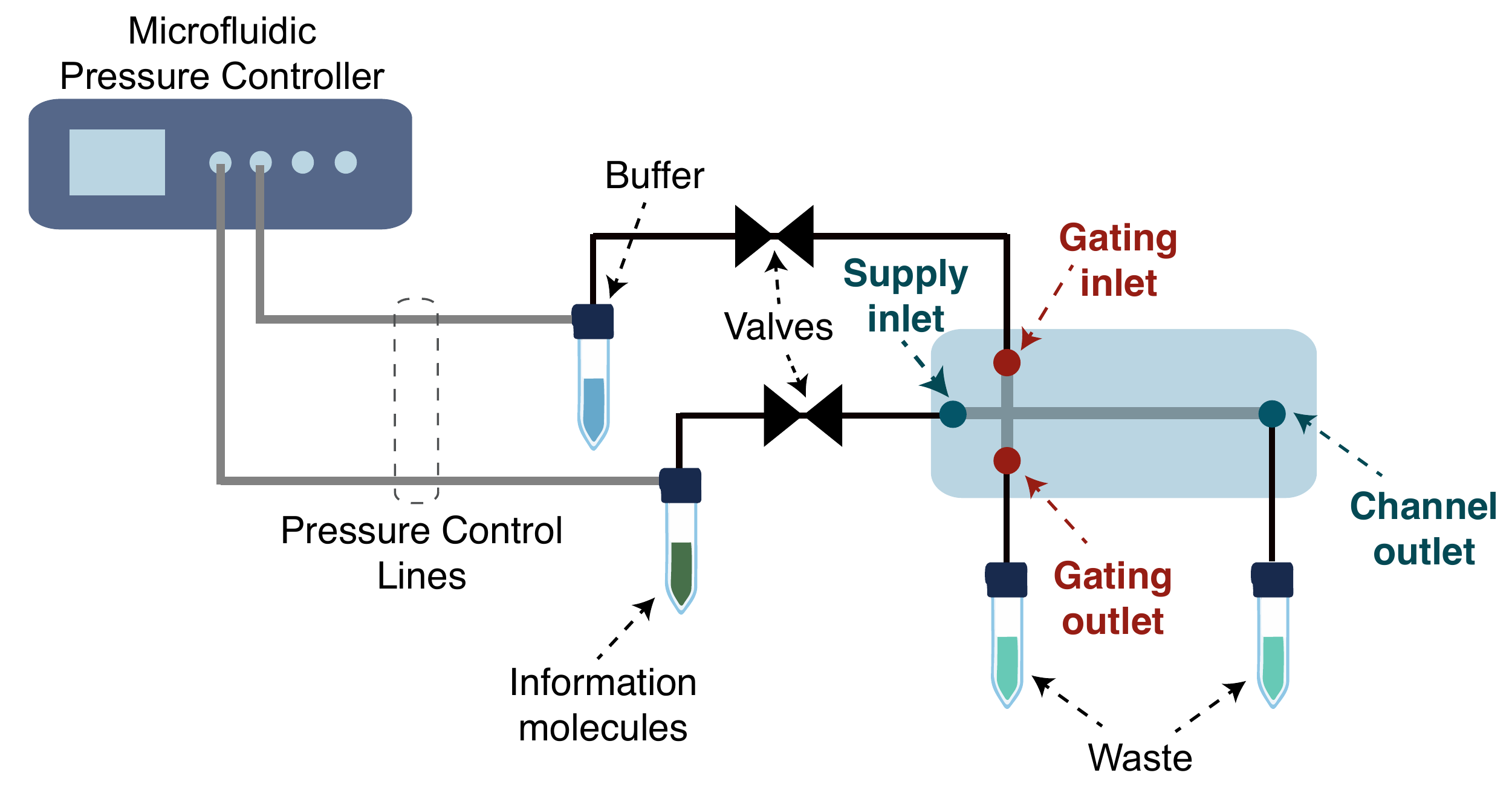}
        \caption{Schematic representation of microfluidic system setup for hydrodynamic gating-based concentration waveform generation.}
    \label{fig:hydrodynamic}
\end{figure}

\begin{figure*}[t!]
    \centering
     \includegraphics[width=0.8\textwidth]{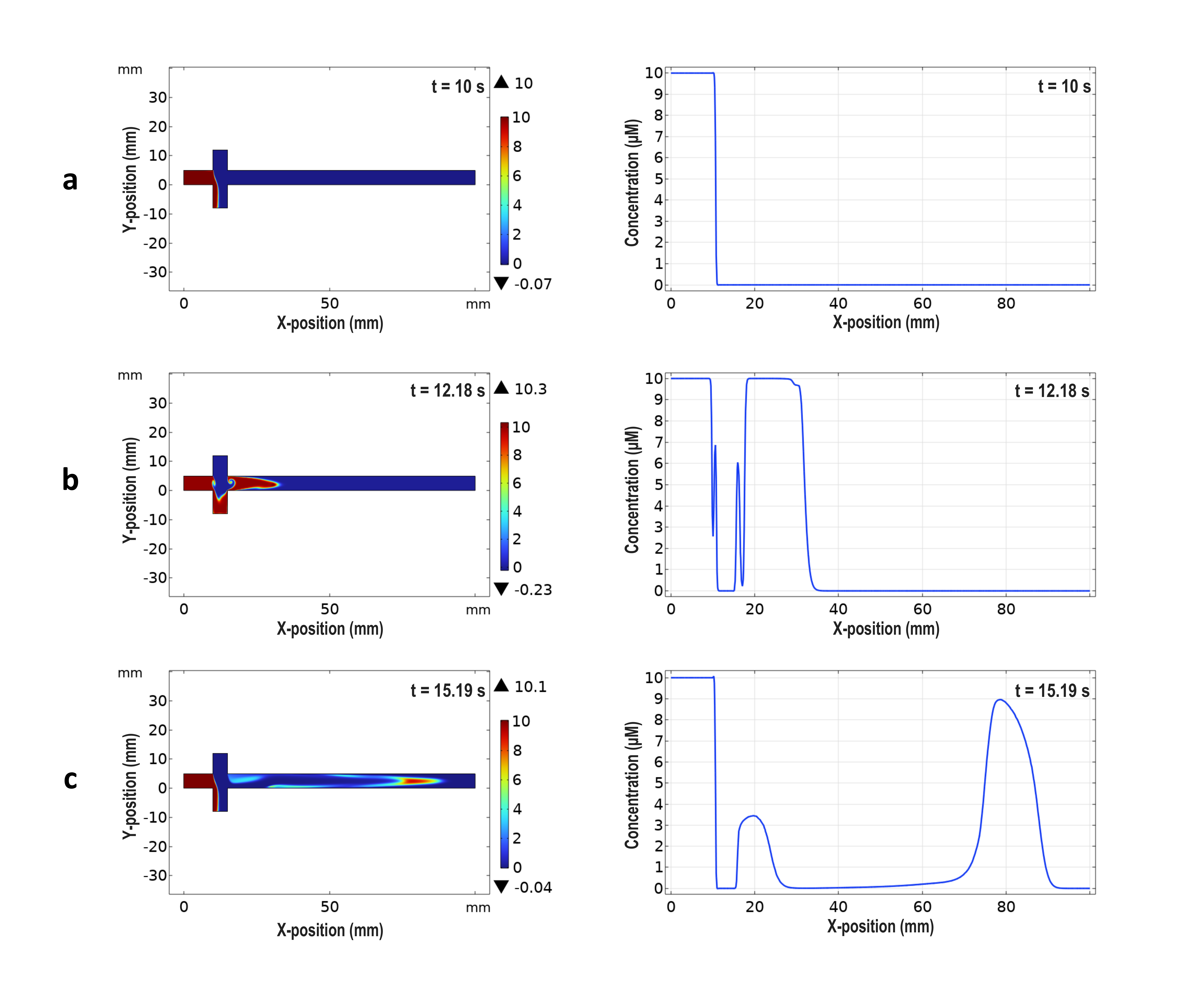}
        \caption{An illustration of a concentration pulse sampled at three different time instances: (a) t = 10.00s, (b) t = 12.18s, and (c) t = 15.19s, corresponding to different states of the hydrodynamic gating process. (a) and (c) depict the gating state, and (b) depicts the injection state. Simulated with COMSOL Multiphysics.}
    \label{fig:hydrodynamic2}
\end{figure*}

Hydrodynamic gating is one of the most widely used hydrodynamic waveform generation techniques, having the potential to generate a rich set of dynamic chemical concentration patterns in microfluidic channels. This method has a high degree of precision while being easily implemented and operated \cite{guo2021time,chen2013analysis}. An example microfluidic setup for the implementation of hydrodynamic gating is conceptually illustrated in Fig. \ref{fig:hydrodynamic}, where the cross-shaped microfluidic channel can be fabricated via soft-lithography techniques using polydimethylsiloxane (PDMS) polymer, and the pressure control at the micro-channel inlets can be achieved by custom-built or commercial pressure control systems, such as programmable syringes. Accordingly, in this setup, hydrodynamic gating can be achieved by switching on and off the laminar flow that comes from the supply inlet and transports the molecules of interest through the on-off modulated laminar flow that is supplied from the gating inlet. When turned on, the flow from the gating inlet generates a hydrodynamic force in the vertical direction across the cross-section, that is sufficient enough to cut the flow from the supply inlet. Therefore, during the on-state of the hydrodynamic gating, the molecules transported through the continuous flow from the supply inlet are not able to enter the microfluidic channel and are directed towards the gating outlet. On the other hand, when gating is turned off for a short duration, the vertical hydrodynamic force is removed, and thus, the molecules coming from the supply inlet are injected into the microfluidic channel. The gating duration is therefore critical for tuning the width of the pulses propagating inside the channel. The other design parameters that play critical roles in determining the shape of generated chemical waveforms include the absolute and relative flow velocities or pressures applied at the supply and gating inlets, and the concentration of molecules transported from the supply inlet \cite{chen2010hydrodynamic}.
As the hydrodynamic gating method involves a relatively small number of design parameters that can be fine-tuned by the state-of-the-art micro/nanofabrication techniques and pressure control devices, it provides a high level of reproducibility, which is crucial for reliable and communicable experiments. The technique has been experimentally shown to generate various chemical pulses and pulse sequences with different frequencies, amplitudes, and shapes \cite{guo2021time, chen2010hydrodynamic, chen2017microfluidic, xue2020microfluidic}.

To validate the high spatiotemporal control promised by the hydrodynamic gating method, we carried out finite element simulations in COMSOL Multiphysics, the results of which are provided in Fig. \ref{fig:hydrodynamic2}. On the left-hand side of the figure is shown the 2D concentration profile of molecules sampled at specific times; and on the right-hand side, the graphs plot the concentration of molecules sampled across the overall microfluidic channel from the supply inlet to the channel outlet using a concentration probe right at the middle of the channel. Accordingly, the concentration profiles observed during the three basic states are provided; (a): the first gating state, (b): the injection state, and (c): the second gating state. During the first gating state (a), all the molecules coming from the supply inlet are diverted to the gating outlet by the vertical flow applied from the gating inlet such that they are not able to pass through the cross-section towards the microfluidic channel. During the injection state, however, the gating flow is turned off for a short time duration such that a small number of molecules can propagate into the microfluidic channel without being diverted. When the gating flow is turned on again, the system returns to its initial state, such that the molecule transport into the microfluidic channel stops. As a result, a short concentration pulse of molecules is generated, which then continues its propagation towards the channel outlet. As the concentration pulse propagates along the microfluidic channel, it experiences dispersion due to the diffusional transport that accompanies conventional transport. The interplay between diffusion and convection determines pulse dispersion, which can be an important metric that is directly connected to the ISI in MC applications. 

\begin{figure*}[t!]
   \centering
   \includegraphics[width=0.8\textwidth]{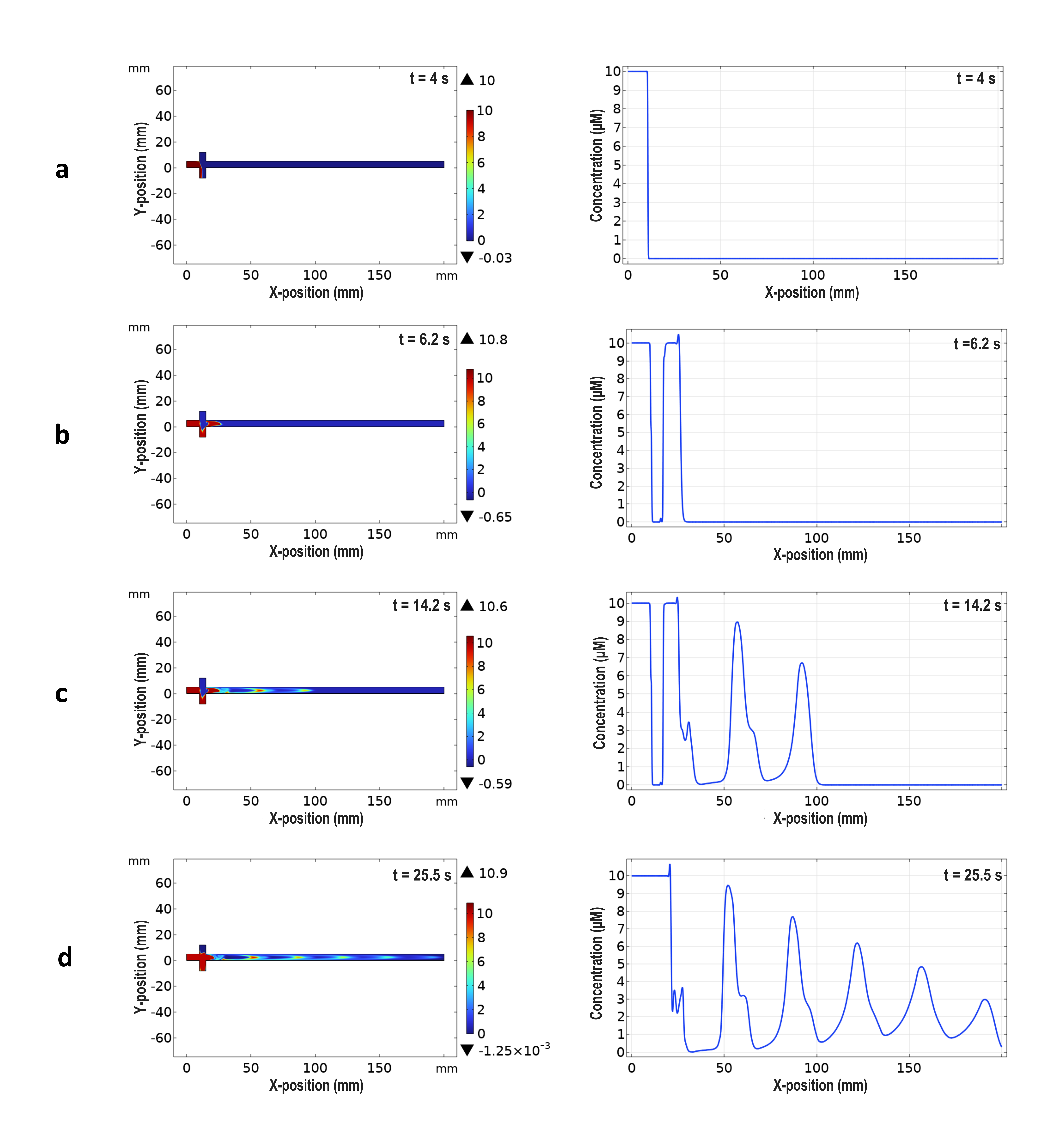}
        \caption{An illustration of consecutive concentration pulses sampled at four different time instances: (a) t = 4.00s, (b) t = 6.20s, (c) t = 14.20s and (d) t = 25.50s. By applying a gating flow from the upper inlet at periodic time instances, consecutive concentration pulses are generated and propagated within the MC channel. Simulated with COMSOL Multiphysics.}
 \label{fig:hydrodynamic3}
\end{figure*}

\begin{figure*}[t!]
   \centering
   \includegraphics[width=0.8\textwidth]{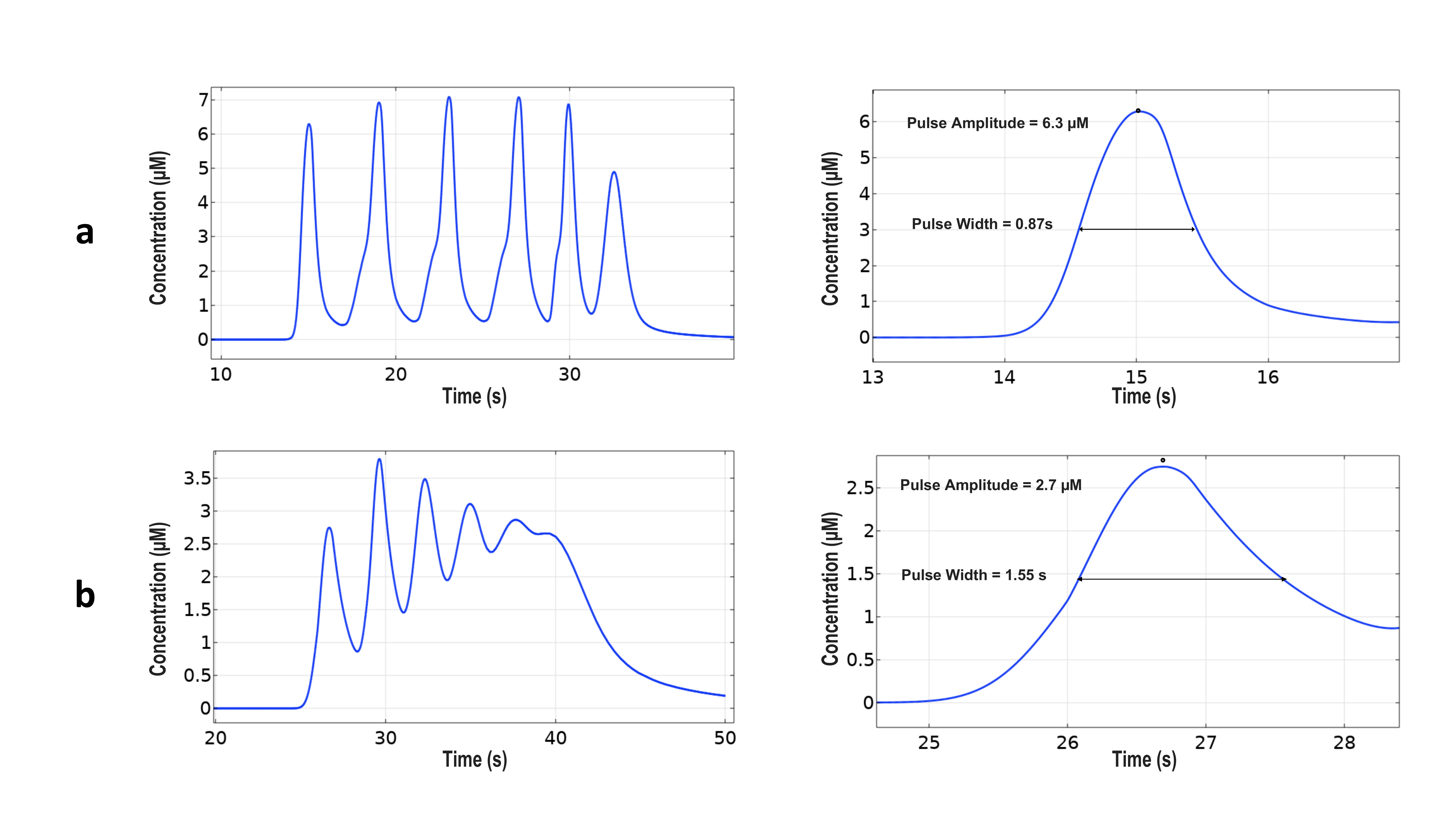}
        \caption{An illustration of the concentration profile of consecutive pulses as a function of time, sampled at periodic time instances at two different points of the channel: (a) The channel's midpoint (b) The channel's outlet. The left side of the figure shows the propagation of all of the consecutive pulses as a function of time, while the results on the right side of the figure illustrate only the propagation profile of the first transmitted pulse as it passes the midpoint and the outlet of the channel. These simulations were carried out using COMSOL Multiphysics.}
 \label{fig:hydrodynamic4}
\end{figure*}

To observe the ISI, we also performed simulations for the consecutive generation of short concentration pulses through the hydrodynamic gating technique. Fig. \ref{fig:hydrodynamic3} shows the 2D and 1D propagation profiles of the five consecutive concentration pulses generated with the same gating durations and even pulse generation intervals. The effect of dispersion on the width and the peak amplitude of the individual concentration pulses can be observed. Although the pulse peaks are easily distinguishable for this setup with a particular channel length, it can be observed that the interference of consecutive pulses increases as they near the end of the microfluidic channel. These results highlight the importance of generating short concentration pulses to avoid or minimize the ISI. 

Fig. \ref{fig:hydrodynamic4} presents the concentration of consecutive pulses over time, measured at two different locations in the channel: the midpoint and the outlet, for better demonstrating the effect of dispersion on the peak amplitude and width. The concentration profile of all of the pulses is shown on the left side of the figure, while the concentration profile of only the first pulse is shown on the right side of the figure, at both the midpoint and outlet of the channel. The examination of the amplitude and the width of the first transmitted pulse demonstrates that the dispersion effect becomes more prominent at the outlet of the channel. It can also be observed that the peak of the last propagated pulse is lower than that of the other five pulses. The reason for this phenomenon is that the velocity within the microchannel decreases significantly after the last gating and pulse creation, causing the last generated pulse that is still propagating in the channel to move slowly towards the outlet and with more dispersion. 

\begin{figure}[t!]
    \centering
    \includegraphics[width=0.48\textwidth]{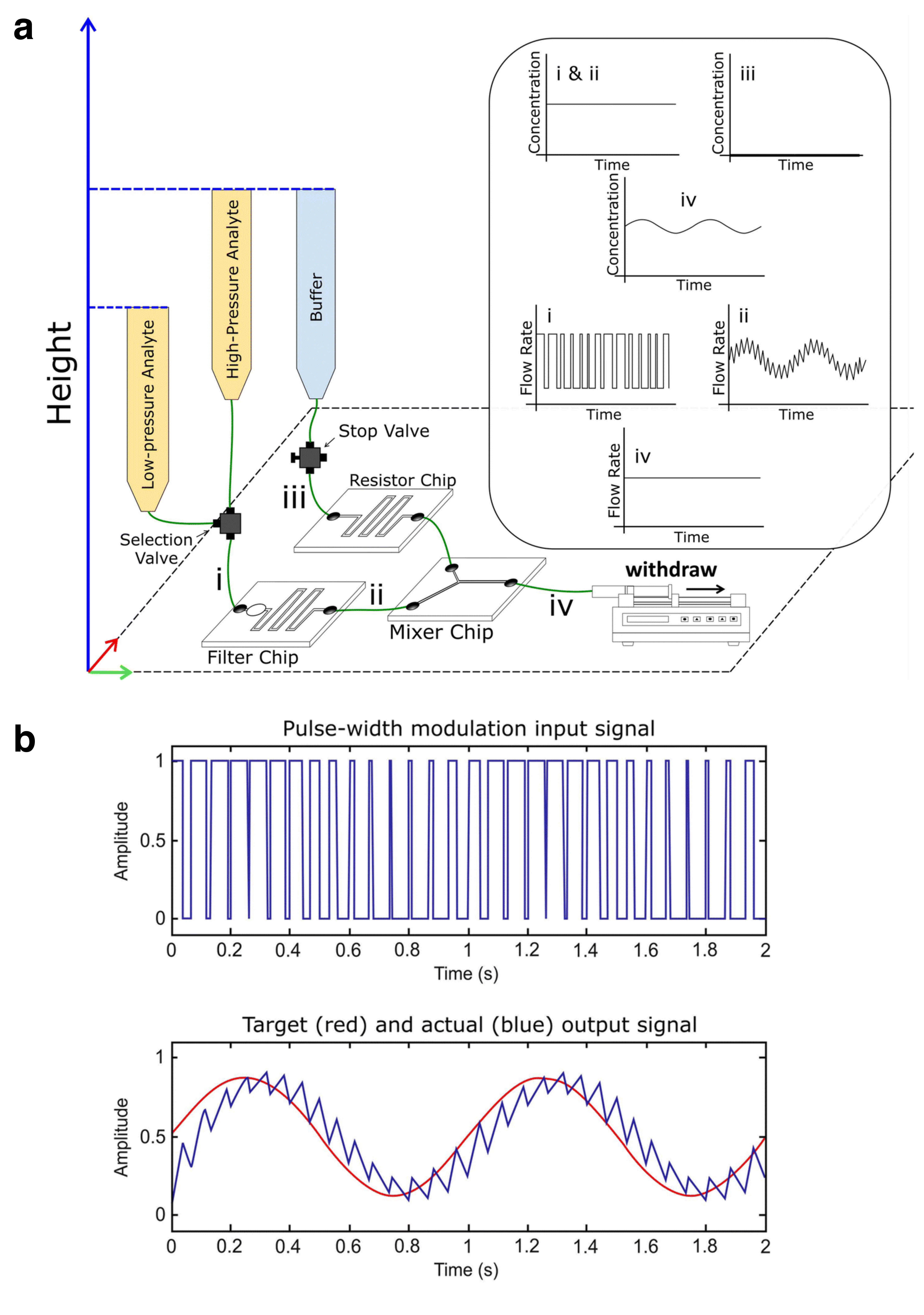}
        \caption{ A programmable microfluidic device designed to generate chemical waveforms hydrodynamically.
        (b) The top plot is the input PWM signal with Y-axis showing the flow rate and the bottom plot is the output signal (ragged sinusoidal) with Y-axis showing the concentration. Reproduced with permission from \cite{garrison2018electrically}.}
         
    \label{fig:programmablehydrodynamics}
\end{figure}

The programmability of hydrodynamic concentration waveform generation methods can be extended through electrical control with the utilization of electrical microfluidic valves with a low response time, which modulate the injection rate of buffer and chemical solutions. By utilizing concepts and tools from electrical engineering and fluid mechanics, this microfluidic system can deliver time-varying concentrations and arbitrary waveforms fast and accurately. In this system, concentration waveforms are modulated through pulse width modulation (PWM), a standard approach for creating analog signals from digital inputs.

An example microfluidic platform combining hydrodynamic gating and electrical control was implemented in \cite{garrison2018electrically}. As shown in Fig. \ref{fig:programmablehydrodynamics}(a), the proposed system is comprised of three different functional microfluidic chips connected to each other: (i) filter chip, (ii) resistor chip, and (iii) mixer chip. Filter chip consists of an elastic membrane-capped cavity that functions as a microfluidic capacitor and a serpentine channel that functions as a microfluidic resistor. The resistor chip contains a serpentine channel whereas the mixer chip has a Y-shaped channel. There are three reservoirs in the system. One of the reservoirs holds the buffer, which is deionized (DI) water, whereas the other two hold fluorescein solutions. These two reservoirs are connected to the inlets of the filter chip through a selection valve and located at different heights to create different hydrostatic pressures. Even though the concentrations of these two solutions are the same, the instantaneous output flow rates are different when the flow selection valve is regulated to switch between them, resulting in differing volumes of the solution flowing into the filter chip per unit time. 

The filter chip uses a combination of a capacitor and a resistor to attenuate the high-frequency components of the PWM signal. Through an elastic membrane-capped cavity, the capacitor allows low-frequency signals to pass through while blocking high-frequency signals, whereas the serpentine channel resistor provides resistance to the flow of current and determines the rate at which the filter circuit responds to changes in the input signal. The filter produces an analog output signal that corresponds to the average pulse period of the PWM signal, and the specific response characteristics of the filter, such as its time constant and cutoff frequency, can be adjusted by changing the design parameters of the capacitor and resistor. 
A reservoir containing the buffer is connected to the inlet of the resistor chip through a stop valve, enabling the solution to be switched off manually. The mixer chip is used to achieve the mixing of the solution with the buffer to generate the final form of the concentration waveform that propagates in the microfluidic channel. By connecting a syringe pump to the mixer's outlet instead of the system inlets, liquid can be withdrawn at a constant flow rate. It should be noted that this configuration is particularly useful in complex microfluidic systems such as this setup where maintaining stable and same flow conditions among multiple interconnected chips is important. 
Using this setup, as shown in Fig. \ref{fig:programmablehydrodynamics}(b), a PWM signal (top plot) is converted into the target signal, i.e., a red sinusoidal wave in the bottom plot. Additionally, by low-pass filtering of the PWM signal, the actual signal (blue ragged sinusoidal wave in the bottom plot) is derived, which closely resembles the red sinusoidal signal of the target. The employment of the electrically controlled PWM technique for the microfluidic device gives the system the ability to rapidly and in a programmable manner generate sinusoidal, triangle, sawtooth, square, and even more complex waveforms with a high level of accuracy, hence, is promising for pulse shaping in microfluidic MC systems \cite{garrison2018electrically}. 


\begin{figure}[t!]
    \centering
    \includegraphics[width=0.48\textwidth]{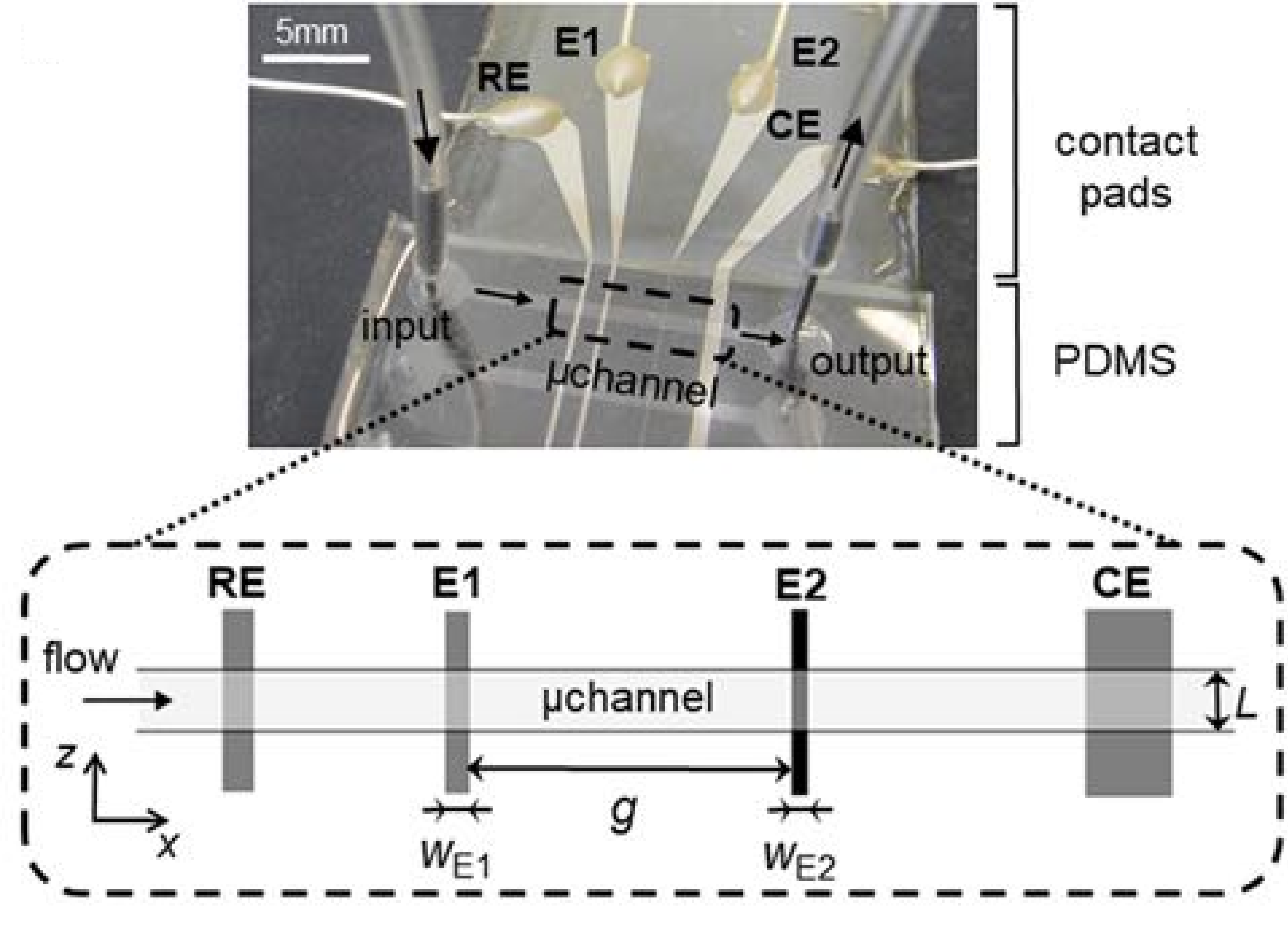}
        \caption{A typical microfluidic device for electrochemical generation of chemical waveforms, consisting of a reference electrode (RE), two working electrodes (E1, E2), and a counter electrode (CE) placed at the bottom of a rectangular microfluidic channel. Reproduced with permission from \cite{abadie2019electrochemical}.}
    \label{fig:electrochemicalmicrofluidcdevice}
\end{figure}

\subsection{Electrochemical Methods}
\label{sec:electrochemical}
Chemical concentration waveforms can also be formed through electrical control by exploiting electrochemical reactions, e.g., reduction-oxidation (redox) reactions, of the propagating molecules on the surfaces of electrodes that are placed inside microfluidic channels. Accordingly, in this method, by positioning the electrodes in various arrangements, concentration gradients can be generated along the microfluidic channel through the interplay between reaction, diffusion, and convection processes. The shape of the resulting concentration gradients or waveforms can also be controlled by space- and time-modulating the electrical potential applied to the electrodes, which modulates the rates of the electrochemical reactions \cite{krabbenborg2014electrochemically}.
The typical implementation of this technique involves a microfluidic device composed of four electrodes, i.e., two working electrodes (E1, E2), a reference electrode (RF), and a counter electrode (CE), all positioned at the bottom of the rectangular microfluidic channel in a dual-channel-electrode configuration, as shown in Fig. \ref{fig:electrochemicalmicrofluidcdevice} \cite{abadie2019electrochemical}. In this example setup, the pseudo-reference electrode is located in the upstream part of the channel to keep the base electrical potential stable during the operation. The microfluidic channel is filled with a solution of electrically neutral molecules, which is continuously supplied from the channel inlet with a constant flow velocity. When the electrical potential of the first working electrode E1 is turned on, the neutral molecules, as they are transported over the electrode through convection and diffusion, undergo an oxidation reaction which converts them into electroactive molecules. Counter electrode, meanwhile, serves as a site for the reduction of the oxidized molecules that are generated at the working electrode, allowing the overall reaction to proceed smoothly and efficiently \cite{delahaye2021electrochemical}. The generated electroactive molecules propagate downstream along the microfluidic channel with a concentration waveform that depends on the duration and the amount of electrical potential applied to E1 in addition to the flow velocity in the channel, and the diffusion coefficient of the molecules. Hence the electrical potential applied to E1 can solely modulate the generated concentration waveforms when the other system parameters are kept constant. The role of the second electrode E2 is to monitor the generated waveforms as they propagate through the channel, and convert them back to neutral molecules through a reduction reaction. Thus, the overall system with two working electrodes can be considered as operating in a generator-collector mode. It is to be noted that the potential difference between E1 and E2 electrodes generates an electric field along the microfluidic channel, which contributes to the transport of molecules downstream in addition to the hydrodynamic forces.

\begin{figure}[t!]
    \centering
    \includegraphics[width=0.45\textwidth]{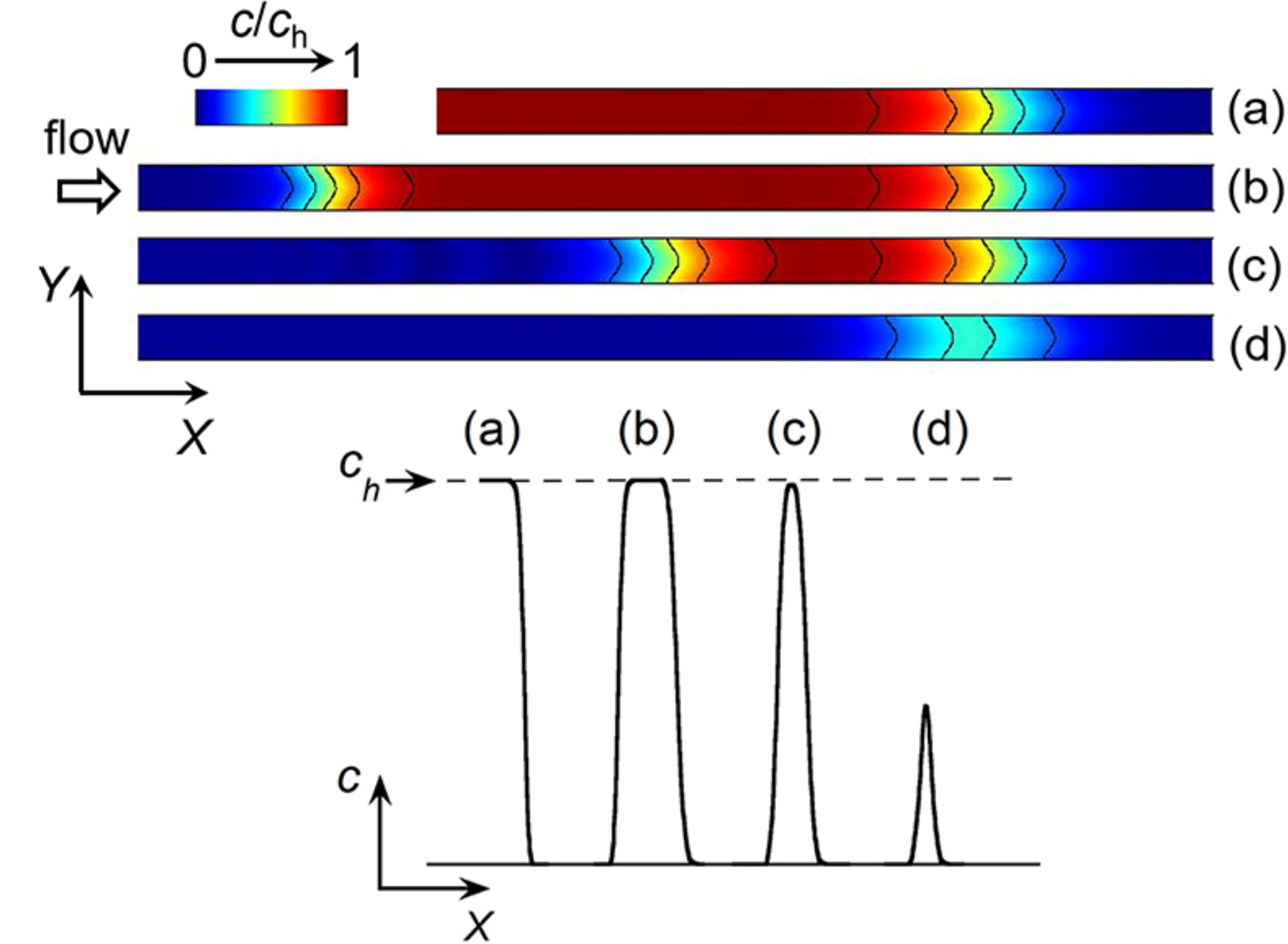}
        \caption{(Top) Chemical concentration signals generated by the electrochemical method, sampled at a given time between E1 and E2 electrodes in the microfluidic device. (Bottom) Corresponding concentration waveforms along the x-axis. Reproduced with permission from \cite{abadie2019electrochemical}.}
    \label{fig:chemicalwaveformselectrochemical}
\end{figure}

The electrochemical method has been theoretically and experimentally shown to generate a wide range of concentration waveforms through the modulation of the first working electrode potential \cite{abadie2019electrochemical, perrodin2020electrochemical}. According to the simulation results shown in Fig. \ref{fig:chemicalwaveformselectrochemical}, the generated waveforms can take various shapes, such as peaks and plugs, depending on the duration of the applied electrical potential. The top section of the figure demonstrates the 2D concentration distribution of four distinct shapes of concentration pulses that can be generated between electrodes E1 and E2 (as shown in the bottom section of the figure). These results were obtained through finite element simulations, sampled at specific time instances. By adjusting the duration and amount of electrical potential applied to E1 and changing the flow velocity in the microfluidic channel, it is possible to generate these unique concentration pulse shapes. The use of electrochemical reactions in microfluidic MC testbeds is promising for pulse shaping. The technique provides a high level of programmability through electrical access and can be miniaturized easily to be integrated into MC testbeds of various scales \cite{abadie2019electrochemical}.

\subsection{Acoustofluidic Methods}
\label{sec:acoustofluidics}
Acoustofluidics, the manipulation of molecules, nanoparticles, and biological entities through acoustic forces in microfluidic structures has been of great interest recently due to its noninvasiveness and ease of integration into microfluidic systems \cite{bruus2012acoustofluidics,rufo2022acoustofluidics}. The acoustic vibration forces that are generated at ultrasonic frequencies in the hundreds of kHz to tens of MHz range have wavelengths that are well-suited to microfluidic channel scales \cite{friend2011microscale}. This method is primarily used in the separation of nanoparticles, cancer cells, bacteria, extracellular vesicles, blood components, droplets, and other particles, which is a fundamental process in bioanalytical research \cite{wu2019acoustofluidic}. The acoustofluidic method has the capability to generate spatiotemporally modulated concentration waveforms in microfluidic channels with the help of a micromixer structure exposed to the acoustic field. Micromixer is typically a microbubble that is trapped inside the microfluidic channel and oscillates at certain frequencies under the effect of an acoustic field. The oscillation of microbubble transduces acoustic force into mechanical force, which in turn, helps rapid mixing of buffer and solution that transports the molecules of interest from the inlet to the microfluidic channel \cite{ahmed2013tunable, destgeer2013continuous}. 

\begin{figure}[t!]
    \centering
    \includegraphics[width=0.45\textwidth]{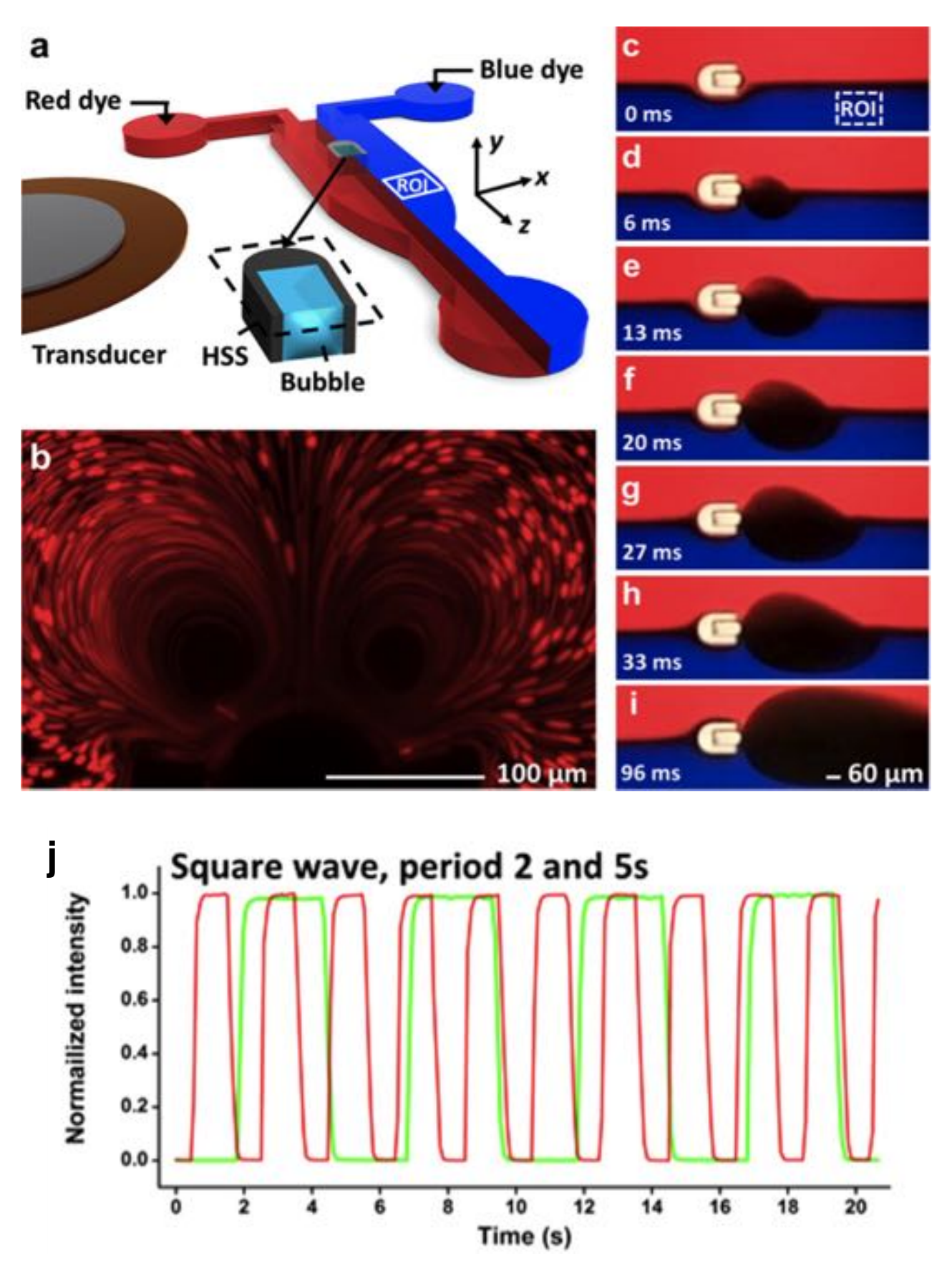}
        \caption{Acoustofluidic chemical waveform generation: (a) Schematic of a typical setup with HSS. Piezoelectric transducers, which are placed adjacent to the microfluidic channel on a glass slide, produce low-intensity acoustic waves. The generated acoustic waves oscillate the bubble trapped in the HSS, which is positioned at the interface of the two liquids. (b) Acoustic microstreaming and flow recirculation during the bubble oscillation. (c-i) Acoustofluidic generation of a chemical waveform observed at different time instances. When an ink solution and a buffer are used as the inlet solutions, the resulting chemical waveforms are monitored through the optical density in the region of interest (ROI). (j) The diagram of the chemical waveforms produced by acoustic signals in the shape of a square wave. Reproduced with permission from \cite{ahmed2014acoustofluidic}.}
    \label{fig:acousticmethod}
\end{figure}

In \cite{ahmed2014acoustofluidic}, Ahmed et al. proposed an acoustofluidic chemical waveform generator based on the active mixing of a buffer with a chemical solution of interest using acoustically driven oscillating bubbles. A schematic diagram of this platform is shown in Fig. \ref{fig:acousticmethod}(a), which exploits the bubble oscillation in an acoustic field to generate arbitrary chemical waveforms inside the microfluidic channel \cite{glynne2012acoustofluidics}. The proposed device consists of a single-layer PDMS-based microfluidic channel with two inlets and one outlet. The architecture of the device, including the number of inlets and outlets, however, can vary depending on its purpose. In this setup, the microfluidic channel is equipped with a horseshoe structure (HSS) that traps a single bubble via surface tension. A piezoelectric transducer, that generates the acoustic field, is placed adjacent to the microfluidic channel. The membrane of the trapped bubble oscillates under an acoustic force field generated by a piezoelectric transducer, which is controlled through an electronic function generator. Maximum bubble oscillation occurs at its resonance frequency, which depends on the bubble size. A pressure gradient is generated in the fluid due to the second-order effect of nonlinearity in the Navier-Stokes equation at the resonance frequency, driving acoustic microstreams. When the trapped bubble is vibrated, the counter-rotating vortices resulting from microstreaming disrupt the smooth buffer-solution interface that is resulting from the laminar flow regime in the microchannel. Through the disruption of the interface, the microstreams drastically enhance the mass transport along the direction perpendicular to the laminar flow, effectively mixing the liquids. This is referred to as the ON state of mixing. This mixing process is shown in Fig. \ref{fig:acousticmethod}(c-i). When the acoustic force field is turned off for a short duration by the piezoelectric transducer, the mixing of inlet solutions stops, and the characteristic laminar flow manifests again. This is referred to as the OFF state of mixing. Hydrodynamic forces dominate the system once the acoustic field is eliminated in the OFF state, and this leads to the propagation of the generated chemical waveforms along the channel. As a result of the fast responses of the electrical and acoustic fields, the system can toggle between ON and OFF states swiftly and transforms the electrical signals into chemical waveforms which will propagate through the channel, as shown in Fig. \ref{fig:acousticmethod}(j)  \cite{ahmed2013tunable, glynne2012acoustofluidics, destgeer2014adjustable}.

\section{Discussion on Pulse Shaping Performance} 
\label{sec:discussion}
This section examines and the performance of the microfluidic chemical waveform generation methods introduced in Section \ref{sec:microfluidic_pulse_shaping} in terms of their capacity to be utilized in microfluidic MC testbeds and systems for pulse shaping. This capacity is evaluated based on a couple of key performance criteria, including spatiotemporal resolution, complexity, repeatability, selectivity, and control over propagation. A summary comparison of the methods based on these criteria is presented in Table \ref{table:comparison}. 

\subsection{Spatiotemporal Resolution}
\label{sec:spatiotemporal_resolution}
As in all communication systems, the achievable communication rate is a critical performance metric for MC systems. Communication rate in MC depends on the concentration pulse generation rate on the transmitter side, and the imperfections in the channel that could lead to the distortion and the dispersion of the generated pulses as they propagate along the channel, which in turn might result in, for example, the ISI. To be able to test a wide range of communication rates in practical MC testbeds with particular channel conditions and receiver designs, the transmitter should be capable of generating concentration pulses with well-defined and persistent waveforms at a high rate. This capability of the transmitter can be evaluated in terms of its temporal resolution. The spatial resolution, on the other hand, measures the capability of the transmitter, i.e., the waveform generation technique, to produce well-defined square concentration waveforms. In other words, it gives a measure of the bandwidth of the generated concentration waveforms. Higher the bandwidth, the wider range of concentration waveforms that the method is able to generate inside the microfluidic channel. The temporal and spatial resolutions are connected to each other in their dependence on the design and operating mechanism of the waveform generation technique, and thus, evaluated in this section together by being referred to as the spatiotemporal resolution.

The spatiotemporal resolution in hydrodynamic and electrochemical methods is at a medium level. 
In \cite{chen2010hydrodynamic}, $20$ consecutive concentration pulses generated via hydrodynamic method could be injected into the microfluidic channel with a time interval of $500$ ms between every two consecutive pulses, demonstrating that the hydrodynamic method is capable of generating chemical waveforms with a temporal resolution of approximately $2$ Hz. Nevertheless, because of the band limitations of a microfluidic channel, distortions are inevitably generated as the pulses propagate along the channel, leading to the degradation of the spatial resolution of this technique. Likewise, the reported temporal resolutions for the electrochemical technique range between $2$-$4$ Hz \cite{abadie2019electrochemical}. Moreover, the pulses generated by these methods well approximate a square-wave right after their generation (see Fig. \ref{fig:hydrodynamic2} and Fig. \ref{fig:chemicalwaveformselectrochemical}), overall indicating a medium level of spatiotemporal resolution. 

In comparison with the other methods mentioned above, acoustofluidic methods were reported to exhibit a higher level of spatiotemporal resolution. The experimental results from \cite{ahmed2014acoustofluidic} reveal that this method provides very high temporal resolution reaching up to $30$ Hz frequency. Also, generated pulses approximate square waves much better compared to other techniques (see Fig. \ref{fig:acousticmethod}), overall making the acoustofluidic methods most promising for the microfluidic MC testbeds supporting high data transmission rates.

\begin{table*}[t!]
\centering
\caption{A Comparison of the Microfluidic Pulse Shaping Methods Based on \\ Key Performance Criteria}
\label{table:comparison}
\scalebox{1.2}{
\begin{tabular}{|c|c|c|c|c|c|}
\hline
\textbf{Methods}                                                                   & \textbf{\begin{tabular}[c]{@{}c@{}}Spatio-temporal  \\      Resolution\end{tabular}} & \textbf{\begin{tabular}[c]{@{}c@{}}Control over \\ Propagation\end{tabular}} & \textbf{Repeatability} & \textbf{Selectivity} & \textbf{\begin{tabular}[c]{@{}c@{}}System \\      Complexity\end{tabular}} \\ \hline
\textbf{\begin{tabular}[c]{@{}c@{}}Hydrodynamic \\ Methods\end{tabular}}     & Medium                                                                               & Medium                                                                       & High                   & Low                  & Low                                                                        \\ \hline
\textbf{\begin{tabular}[c]{@{}c@{}}Electrochemical \\ Methods\end{tabular}} & Medium                                                                               & High                                                                         & Medium                 & High                 & Medium                                                                     \\ \hline
\textbf{\begin{tabular}[c]{@{}c@{}}Acoustofluidic \\ Methods\end{tabular}}   & High                                                                                 & Low                                                                          & High                   & Low                  & High                                                                       \\ \hline
\end{tabular}}

\end{table*}

\subsection{Control over Propagation} 
\label{sec:control_over_propagation}
In microfluidic MC testbeds, not only the generation of arbitrary chemical waveforms but also their propagation and the evolution of their profile along the microfluidic channels can be important to accurately test the performances of MC modulation and detection techniques and validate MC channel models. The assessment for control over the propagation of each method considers to what extent the technique that generates the pulses is also able to control their propagation inside the channel, independently of the generation process.

Among the chemical waveform generation methods considered in this paper, the electrochemical method provides the highest degree of control over the propagation of generated waveforms. In the electrochemical method, the redox reactions between the molecules and the first working electrode convert the electrically neutral molecules into electroactive molecules, meaning that they can be influenced by the electric field effect. As there are two working electrodes (E1 and E2, see Fig. \ref{fig:electrochemicalmicrofluidcdevice}), separated by a certain distance in the microfluidic channel, the potential difference between them generates an electric field along the channel, which acts upon the generated concentration signal of the electroactive molecules. Therefore, depending on the polarity of the electroactive molecules, and the magnitude of the potential difference between electrodes, the propagation characteristics of the generated concentration waveforms can be modulated independently of their generation process \cite{krabbenborg2014electrochemically}.

In the hydrodynamic method, however, both the generation and the propagation of concentration waveforms are controlled by hydrodynamic forces. This limits the ability to control propagation characteristics independently of the waveform generation process inside the channel. Thus, the hydrodynamic technique can be considered to have a medium level of control over the propagation of generated waveforms in comparison to the other two methods \cite{chen2010hydrodynamic}.

On the other hand, the acoustofluidic method has the lowest level of control over the propagation of the generated waveforms, as the bubble oscillations due to the acoustic force fields create space-limited vortices, localized only around the bubble contained in HSS. Therefore, the resulting vibrational force fields do not extend into the microfluidic channel, limiting the capacity of this method to dynamically tune the propagation characteristics of the generated chemical waveforms. This lack of control over propagation in this method, however, can be improved with the modulation of the hydrodynamic forces inside the channel, for example, by tuning the inlet pressures \cite{ahmed2014acoustofluidic, destgeer2014adjustable}.

\subsection{Repeatability}
\label{sec:repeatability}

The repeatability of the chemical waveform generation techniques can be evaluated based on the number of successively generated waveforms that have approximately the same shape when introduced into the microfluidic channel. The repeatability is of crucial importance for practical MC testbeds and systems, as MC scenarios typically require the transmission of multiple concentration pulses successively, for example, to evaluate the impact of ISI, and determine the achievable data transmission rates. 

An important factor in generating approximately the same waveforms, and therefore having high repeatability in each transmission is the stability and durability of the microfluidic chip structure, especially in the transmitter part. As is shown in Fig. \ref{fig:hydrodynamic3} and Fig. \ref{fig:acousticmethod}, consecutively generated pulses by both hydrodynamic and acoustofluidic methods have approximately the same waveform, indicating the superb repeatability of these methods. This is due to the fact that all the parameters involved in the design of the transmitter structure and the microfluidic chip for both acoustic and hydrodynamic based methods can be controlled and engineered to yield utmost stability in generating concentration waveforms \cite{chen2010hydrodynamic, ahmed2014acoustofluidic}.

The electrochemical method, on the other hand, can be considered to have lower repeatability over long operation times, due to the chemical reactions involved in the generation of the waveforms. More specifically, the chemical reactions on the electrode surface are highly prone to the effects of environmental fluctuations, such as those in the temperature. Second, chemical reactions are inherently stochastic processes, adding to the low-level repeatability of the electrochemical method.

\subsection{Selectivity}
\label{sec:selectivity}
In practical MC applications, there could be many different types of molecules co-existing in the MC channel, that can result from a biological process at the background or another MC network accessing the same channel. The co-existence of different types of molecules can lead to substantial interference with the communication process. Researchers have already developed several theoretical detection and channel sensing techniques to reduce or eliminate the effect of such background or multi-user interference \cite{kuscu2019channel}. To enable the performance tests of these methods under interference, and the development of more practical techniques to cope with it, microfluidic MC testbeds should be able to replicate the interference conditions. From the transmitter aspect, the pulse shaping technique should be able to selectively control the waveform generation process, such that only the molecules of interest, among other interfering molecules, are modulated in concentration. The electrochemical waveform generation method can be considered to support selectivity for such MC scenarios, due to the inherent specificity of the chemical reactions involved in the waveform generation process \cite{abadie2019electrochemical}. 

In the hydrodynamic method, however, the concentration of molecules is modulated solely by hydrodynamic forces, which does not make a distinction between different types of molecules co-existing in the channel. The only factor that can enable selectivity in hydrodynamic waveform generation is the diffusion coefficient of molecules, which may differ depending on the size of the molecules. However, as the effect of diffusion on the shape of generated waveforms is limited, manifesting itself only in the dispersion of the waveforms as they propagate, it cannot enable the required level of selectivity in pulse shaping in MC testbeds. 

Similar to the hydrodynamic method, the selectivity of the acoustofluidic methods is low. The external acoustic force fields applied by the transducer in this method modulate the oscillation frequency of the bubbles trapped inside the HSS structure, which then only indirectly affect the transport of the molecules around the structure through the hydrodynamic forces. As there is no direct impact of the acoustic force fields on molecular transport, the method cannot selectively distinguish between different types of molecules. Hence, the only element that can lead to a low level of selectivity in this method could be again the diffusion coefficient that depends on the type, i.e., the size, of molecules. \cite{huang2018sharp,ahmed2014acoustofluidic}.

\subsection{System Complexity}
\label{sec:system_complexity}
The complexity of the microfluidic chemical waveform generation techniques is evaluated based on the number and the heterogeneity of individual components required for the waveform generation, and the fabrication methods by which the overall system, including the microfluidic chip and other external components, is assembled to generate chemical waveforms. Systems with less complexity, of course, can be favored because of the ease and low cost of the fabrication process, and more importantly, the increased level of reproducibility. 

Acoustofluidic waveform generation is typically based on a complex system architecture involving a multitude of external components, such as a piezoelectric transducer capable of producing low-intensity acoustic waves, and a horseshoe structure (HSS), which traps the bubble that is oscillated via the acoustic waves inside the microfluidic channel. 

The electrochemical waveform generation method has a medium level of complexity compared to the other two methods. In terms of fabrication, the system architecture consists of a hybrid PDMS-glass chip. Electrodes can be deposited on the glass substrate with a specified spacing with a sputtering mechanism. The waveform generation process in this method is based on the consecutive electric potential pulses applied to the electrodes and the subsequent electrochemical reactions of the molecules on the first working electrode. Therefore, the electrochemical method can be considered to require a much simpler fabrication process and have a simpler operating mechanism compared to the acoustofluidic method \cite{krabbenborg2014electrochemically, abadie2019electrochemical}. 

Hydrodynamic methods, on the other hand, have a less complex system architecture consisting of only a PDMS-based microfluidic chip at the fundamental level. Note that the microfluidic chip and the external pressure control systems that drive the fluid flow inside the channels are common to all microfluidic waveform generation techniques investigated in this paper. The waveform generation process is also relatively simple in hydrodynamic methods, as it only requires the hydrodynamic forces which are inherent to the microfluidic systems \cite{chen2010hydrodynamic, abadie2019electrochemical}. These make the hydrodynamic method the least complex waveform generation method considering that the other methods require additional external force fields, such as acoustic fields, or electrochemical reactions.

\section{Discussion on Applications and Feasibility} 
\label{sec:discussion-feasibility}

Most of the MC applications envisioned in the literature, including those concerning intrabody environments, can potentially be facilitated by high-resolution and programmable MC pulse shaping techniques integrated into MC transmitter architectures. This is because high-resolution control over the waveform of the generated concentration pulses can help address the problems resulting from ISI, and nonlinear and time-varying channel and transmitter/receiver characteristics, as explained in Section \ref{fig:pulse_shaping}. Many of these problems require the generation of various concentration waveforms, and sometimes the ability to adaptively tune their properties in order to improve the accuracy of information transfer. Understanding the relationship between physical design parameters and generated pulse shapes can reveal practical limitations and identify areas of further engineering and optimization, as well as inform the design of optimal and practical MC modulation and detection techniques. This practical understanding is crucial for moving beyond the commonly utilized assumption of instant, impulsive molecule release with point transmitters, which lacks practical relevance \cite{kuscu2019transmitter}.

Microfluidic pulse shaping techniques discussed in this paper are particularly promising for integration into microfluidic MC testbeds. These techniques can allow for a broader range of MC scenarios to be practically tested and enable more accurate and programmable platforms for optimization and validation. The implementation of these practical MC pulse shaping techniques in microfluidic MC testbeds can immediately make the following contributions to the MC literature:

\begin{itemize}
\item Enabling the accurate testing of existing or new MC modulation and detection methods, and facilitating the design of realistic MC techniques that are compatible with the pulse shapes that can be practically generated.
\item Optimizing transmitted pulse shapes (e.g., the duration of a rectangular concentration pulse) from a communication perspective for different practical MC channel and receiver architectures and configurations, such as MC receivers with ligand receptors that may experience saturation depending on concentration pulse amplitude and duration. Such optimization can be readily validated through the integration of the investigated pulse shaping techniques into microfluidic MC testbeds.
\item Eventually, optimizing the overall MC system, including the physical receiver architecture, under various MC scenarios given the pulse shaping technique and the
corresponding set of concentration waveforms that can be generated.
\end{itemize}

Furthermore, leveraging the chemical waveform generation techniques with MC tools and theories in microfluidic chips can significantly contribute to lab-on-chip and organ-on-chip technologies. For example, replicating the signaling dynamics between cell groups and tissues in microfluidic chips is essential for creating organ function in organ-on-chip systems. High-resolution and programmable microfluidic pulse shaping techniques that allow for the implementation of various MC modulation techniques in microfluidic chips can enable ICT-based organ-on-chip platforms that can accurately control chemical signal patterns between different biological components, providing an opportunity to probe and study the role of information flow in microphysiological systems. Such ICT-based platforms can open new avenues in drug delivery and drug design research. Merging ICT with microfluidic technologies through MC pulse shaping techniques can also have implications in biomedical sensor research. These multi-disciplinary platforms can allow for the information-theoretical optimization of sensor architectures for sensing time-varying concentration waveforms of biomarkers generated through microfluidic MC pulse shaping techniques. 

It should be noted, however, that none of the chemical waveform generation techniques investigated in this paper have been demonstrated in \textit{in vivo} environments, such as inside the human body. This raises the question of whether the microfluidic MC pulse shaping techniques can be integrated into practical MC transmitters for \textit{in vivo} MC applications. The main challenge in implementing these techniques in \textit{in vivo} environments will be system complexity, as discussed in Section \ref{sec:system_complexity}, and the spatiotemporal control of pulse shapes independently of the surrounding biological force fields. For example, the high-level complexity of the acoustofluidic pulse shaping techniques, which necessitate fine tuning of the external acoustic transducer and the horseshoe structure for bubble oscillation, may hinder their feasibility for such applications. On the other hand, while the system complexity of hydrodynamic methods is relatively low, the pulse shaping function in this method is based on tuning hydrodynamic force fields, e.g., fluid flow rate, which may not be feasible to control independently of the flow conditions of biological fluids, e.g., blood. Electrochemical methods, although having advantages in terms of system complexity and scalability, may lead to biocompatibility issues due to the required electrochemical reactions. In summary, the investigated microfluidic MC pulse shaping methods are currently not feasible for integration into \textit{in vivo} MC applications. However, further advances in micro/nanotechnologies may provide new opportunities for improving their practicality or offer alternative pulse shaping methods that are more compatible with such applications.

\section{Conclusion}
\label{sec:conclusion}
This paper provided a comprehensive overview of practical microfluidic chemical waveform generation techniques which are promising as pulse shaping methods for microfluidic MC systems and testbeds. Pulse shaping is essential for accurate and high-rate information transfer in MC systems because the inherent low-pass characteristic of the MC channel  leads to significant dispersion of the molecular signals as they propagate. The programmability of the spatiotemporal distribution of molecules of interest inside microfluidic channels is also crucial for extending the capabilities of microfluidic MC testbeds. The chemical waveform generation techniques highlighted in this paper utilize different external forces, such as hydrodynamic and acoustic force fields, or electrochemical reactions, to program the spatiotemporal distribution of molecules inside the microfluidic channels. These methods were analyzed in terms of their operating mechanisms and the characteristics of the generated concentration signals. To accurately assess the suitability and potential utility of these techniques for application in microfluidic MC systems and testbeds, we identified a set of key performance criteria including spatiotemporal resolution, control over propagation, system complexity, repeatability, selectivity, and compatibility with a wide range of applications. Through this comprehensive evaluation, we aim to bridge the gap between theory and practice in MC technology. We believe that this review and the accompanying evaluation will help researchers incorporate programmable, high-resolution pulse shaping techniques into their microfluidic MC testbeds for a more accurate assessment of the developed MC techniques, such as modulation and detection techniques, with well-defined MC signal waveforms inside microfluidic channels.

\section*{Acknowledgment}
\label{Acknowledgment}
This work was supported in part by The Scientific and Technological Research Council of Turkey (TUBITAK) under Grant \#120E301, and European Union’s Horizon 2020 Research and Innovation Programme through the Marie Skłodowska-Curie Individual Fellowship under Grant Agreement \#101028935.

\bibliographystyle{IEEEtran}
\bibliography{microfluidic}

\end{document}